\documentclass[12pt]{iopart}

\usepackage{graphicx}
\usepackage{color}
\usepackage{iopams}

\begin{document}

\title{Skyrme effective pseudopotential up to next-to-next-to leading order}

\author{D. Davesne}
\address{Universit{\'e} de Lyon, F-69622 Lyon, France, \\
Universit\'e Lyon 1, Villeurbanne;  CNRS/IN2P3, UMR5822, IPNL}

\author{A. Pastore}
\address{Institut d'Astronomie et d'Astrophysique, CP 226, Universit\'e Libre de Bruxelles, B-1050 Bruxelles, Belgium}

\author{J. Navarro}
\address{IFIC (CSIC-Universidad de Valencia), Apartado Postal 22085, E-46.071-Valencia, Spain}

\date{\today}
\begin{abstract}
The explicit  form of the next-to-next-to-leading order (N$^2$LO) of the Skyrme effective pseudopotential compatible with all required symmetries and especially with gauge invariance  is presented in Cartesian basis. It is shown 
in particular that for such a particular pseudopotential there is no spin-orbit contribution and that the D-wave term suggested in the original Skyrme's formulation does not satisfy the  
invariance properties. The six new N$^2$LO terms contribute to both the equation of state and Landau parameters. These contributions to symmetric nuclear matter are explicitly given and discussed.
\end{abstract}

\maketitle

\section{Introduction}

Since the implementation of Vautherin and Brink \cite{vau72}, the Skyrme interaction \cite{sky56,sky59} has become a popular tool for the description of nuclear properties based on a self-consistent mean-field approach. In its standard form it consists of zero-range central, spin-orbit and density-dependent terms, involving up to 11 parameters, which are fitted to properties of infinite nuclear matter and some selected nuclei. 
Systematic studies of binding energies and one-body properties of nuclei can thus be successfully performed in a very wide region of the nuclear chart \cite{ben03}. However, there are some nuclear properties which cannot correctly be reproduced with the standard Skyrme terms, especially as one moves away from the valley of $\beta$-stability \cite{erl12}.  Intense work is currently being devoted to improve the existing parametrizations, in particular considering additional terms to the standard form \cite{rai11}.

In the original proposal, an effective pseudopotential was constructed as an expansion of the nuclear interaction in relative momenta, thus simulating finite-range effects in a zero-range interaction. It actually contained more terms than the current standard form, which is limited to second order in momenta plus a density dependent term which comes from a zero-range three-body force. These additional terms 
have received special attention in the last years, in particular the second order tensor term \cite{les07}, a D-wave term \cite{ben13} as well as some three- and four-body terms \cite{sad12,sad13}. 

As an alternative route, the underlying energy density functional (EDF) kernel has been considered \cite{per04}, with no direct references to the effective interaction neither to the Hartree-Fock approximation. In principle, dealing with functional coupling constants instead of pseudopotential parameters introduces more degrees of freedom in the fitting process. It is nevertheless interesting to keep in mind the interplay between the  EDF and the pseudopotential descriptions \cite{lac09,dug09}. The EDF related to the standard Skyrme interaction contains up to second-order derivatives of nuclear densities. In the search for a universal EDF, a connection with underlying nucleon interactions has been investigated \cite{car08}, aiming at obtaining the general EDF structure guided solely by symmetry principles. An expansion of the EDF kernel has been given including up to sixth-order derivatives, that is up to next-to-next-to-next leading order (N$^3$LO). In an analogous way, it has been shown how to construct the associated pseudopotential up to sixth order in the relative momenta \cite{rai11}. The number of coupling constants and interaction parameters increases dramatically as the order increases. However, imposing gauge invariance introduces specific relations between the different terms that considerably reduce that number. 

 In this work we start from the general N$^{2}$LO  pseudopotential derived in \cite{rai11} and give its explicit 
 form in the more familiar Cartesian basis, constraining it to be gauge invariant. 
Although this symmetry is not explicitly required from basic principles, there is some current discussion about the necessity of imposing it in general, since it has been shown \cite{rai11b} that gauge invariance is equivalent to continuity equation for local potentials. The continuity equation is of particular interest in view of using such a pseudopotential for calculations of the time evolution of a quantal system.
For this reason, and from the fact that a local potential is automatically gauge invariant \cite{doba95,bla86}, we restrict ourselves to pseudopotentials which are gauge invariant at all orders. 

The article is organized as follows. In section \ref{building:n2lo} we write the N$^{2}$LO pseudopotential in Cartesian basis and we discuss some properties related to gauge invariance. In section \ref{inm} we derive the contribution of the new terms to the equation of state and Landau parameters of symmetric nuclear matter. The last section will give a brief summary and draw some conclusions.

\section{Construction of the N$^{2}$LO pseudopotential}\label{building:n2lo}

For consistency we will briefly sketch now the notation and general results of \cite{rai11}.  
The Cartesian representation is usually employed to write the Skyrme interaction. However, the general pseudo-potential is more conveniently constructed in the spherical-tensor representation. With the conventions of \cite{rai11}, the three components of a rank-1 tensor as the Pauli matrices are written as
\begin{eqnarray}
\sigma^{(i)}_{ 1,\mu=\left\{-1,0,1\right\}} &=&
-i \left\{\frac{ 1}{\sqrt{2}}\left(\sigma^{(i)}_{ x}  -i\sigma^{(i)}_{ y}\right),
 \sigma^{(i)}_{ z}, 
\frac{-1}{\sqrt{2}}\left(\sigma^{(i)}_{ x} +i\sigma^{(i)}_{ y}\right)\right\}.
\end{eqnarray}

The mathematical construction of a pseudopotential is equivalent to the construction of scalars
with relative momenta ${\bf k}$, ${\bf k'}$,  and 
$\bsigma^{(1,2)}$ as basic ingredients, with the usual definitions ${\bf k} = ( \overrightarrow{\nabla}_1 - \overrightarrow{\nabla}_2)/2i$ and $ {\bf k}' =  - ( \overleftarrow{\nabla}_1 - \overleftarrow{\nabla}_2)/2i$. 
Thus we have to consider, up to a given order, all possible tensors built with these quantities and couple them to get a scalar.
In the so-called LS-coupling, the general potential reads: 
\begin{equation}
 \hat{V}=\sum_{{\tilde{n}' \tilde{L}', \\ \tilde{n} \tilde{L},v_{12} S}}
      C_{\tilde{n} \tilde{L},v_{12} S}^{\tilde{n}' \tilde{L}'}
\hat{V}_{\tilde{n} \tilde{L},v_{12} S}^{\tilde{n}' \tilde{L}'} \, ,
\label{hatV}
\end{equation}
with 
$C_{\tilde{n} \tilde{L},v_{12} S}^{\tilde{n}' \tilde{L}'}$ the coupling constants and
\begin{eqnarray}
 \nonumber \hat{V}_{\tilde{n} \tilde{L},v_{12} S}^{\tilde{n}' \tilde{L}'}=
&&\frac{1}{2}i^{v_{12}} \left( \left[ \left[K'_{\tilde{n}'\tilde{L}'}
                                            K_ {\tilde{n }\tilde{L}}\right]_{S}
\hat{S}_{v_{12} S }\right]_{0}  
+ (-1)^{v_{12}+S} \left[ \left[K'_{\tilde{n} \tilde{L}}
                                        K_ {\tilde{n}'\tilde{L}'}\right]_{S}
\hat{S}_{v_{12} S}\right]_{0}   \right) \\
&&\times \left(1-\hat{P}^{M}\hat{P}^{\sigma}\hat{P}^{\tau}\right)
\delta({\bf r'}_1-{\bf r}_1) \delta({\bf r'}_2-{\bf r}_2) \delta({\bf r}_1-{\bf r}_2).
\label{VV}
\end{eqnarray}
By construction, this potential is invariant under space rotation, space inversion, time reversal and hermitian conjugation.  
$K_{\tilde{n} \tilde{L}}$ and $K'_{\tilde{n} \tilde{L}}$ are tensors of order $\tilde n$ and rank $\tilde L$ in the relative momenta ${\bf k}$ and ${\bf k'}$, respectively (see \cite{rai11} for explicit expressions). They are coupled to a total angular momentum $S$, which is coupled to the spin part to get a zero total angular momentum.
The spin part is:
\begin{eqnarray}
    \hat{S}_{v_{12} S} =\left(1-\frac{1}{2}\delta_{v_1,v_2}\right)\left(
[\sigma^{(1)}_{v_1}\sigma^{(2)}_{v_2}]_S +
[\sigma^{(1)}_{v_2}\sigma^{(2)}_{v_1}]_S \right),
\end{eqnarray}
where $v_{12}=v_1+v_2$. In the last line of Eq. (\ref{VV}), $\hat{P}^{M}$,  $\hat{P}^{\sigma}$ and  $\hat{P}^{\tau}$ are exchange operators in direct, spin and isospin spaces, respectively, and the sum of indices $\tilde{n}$ and $\tilde{n}'$, which must be even so that the potential is invariant under space inversion, fixes the order of the expansion. 

To obtain the pseudopotential one has to constrain it with general symmetry properties, in particular Galilean invariance in the non-relativistic case.
In doing so we get 2, 7 and 15 independent parameters at 0th, 2nd and 4th order, respectively. 
A possible choice to reduce these numbers consists in imposing some other additional symmetry, namely the gauge invariance. It does not introduce any further simplification at 0th and 2nd order, while at 4th order we are left with only 6 independent parameters. 

Even if gauge invariance is imposed in a different way on the energy functional density and on the pseudo-potential \cite{rai11,car08}, the consequence is always to provide some relations between the different coupling constants so that only specific combinations with $\bf k'$, $\bf k$ and $\bsigma^{(1,2)}$ can survive.
More specifically, in the case under consideration, the $U(1)$-type gauge transformation acts on a wave function as a multiplicative phase
\begin{equation}
|\Psi'\rangle = \exp \left( i \sum_{j=1}^{A}\phi({\bf r}_j) \right)|\Psi\rangle, 
\label{transfo}
\end{equation}
where $\phi({\bf r}_j)$ is an arbitrary real function depending on the position.
To see the practical consequences of this invariance on the pseudopotential, it is convenient to consider a general 2-body potential $V({\bf r}'_1, {\bf r'}_2, {\bf r}_1, {\bf r}_2)$, where for simplicity spin and isospin indices are dropped since they are not relevant for the argument. The interaction energy can then be written as:
\begin{equation}
E = \int d{\bf r}'_1 d{\bf r}'_2 d{\bf r}_1 d{\bf r}_2 V({\bf r}'_1, {\bf r}'_2, {\bf r}_1, {\bf r}_2)  
  \left[ \rho({\bf r}_1, {\bf r}'_1) \rho({\bf r}_2, {\bf r}'_2)  -\rho({\bf r}_2, {\bf r}'_1) \rho({\bf r}_1 {\bf r}'_2)\right] 
\end{equation}
The gauge tranformation on the $A$-body wave function defined above implies that the density matrix transforms as $\rho'({\bf r}, {\bf r}') = \exp(-i [\phi({\bf r})-\phi({\bf r}') ] )\rho({\bf r},{\bf r}')$. 
In the case of a local interaction, the presence of two delta functions
$\delta({\bf r}'_1 - {\bf r}_1)\delta({\bf r}'_2 - {\bf r}_2) $ in the above integral guarantees automatically the gauge invariance \cite{doba95}.  

Concerning the pseudopotential, the transformation (\ref{transfo}) translates into:
\begin{equation}
\hat{V'}= e^{-i\phi({\bf r}'_2)} e^{-i\phi({\bf r}'_1)} \hat{V} e^{i\phi({\bf r}_1)} e^{i\phi({\bf r}_2)}.
\end{equation}
The invariance $\hat{V'}=\hat{V}$ then leads to the following equation 
\begin{equation}
[\phi({\bf r}_1),\hat{V}] + [\phi({\bf r}_2),\hat{V}] = 0,
\label{gauge}
\end{equation}
which has to be imposed order by order because of the velocity-dependent terms entering (\ref{hatV})-(\ref{VV}).

 As mentioned before, up to second order there are only nine independent parameters. As shown in \cite{rai11}, all together define the parameters of the standard Skyrme interaction, including tensor and spin-orbit terms, but excluding density-dependent terms, which are originated from the three- and four-body contributions not considered here. As gauge invariance is automatically satisfied up to second order, the standard interaction leads to a coherent approach.
From the 15 different terms of the general pseudopotential (\ref{hatV}) at
fourth order, we have identified the combinations satisfying  condition (\ref{gauge}) and discarded the remaining non-invariant terms.

As a result, we are left with only  6 independent parameters, thus confirming the findings of \cite{rai11}. We chose them as $C_{22,00}^{22}$, $C_{22,20}^{22}$, $C_{11,00}^{31}$, $C_{11,20}^{31}$, $C_{22,22}^{22}$, and $C_{11,22}^{33}$.  
Afterwards we realized that it is more convenient to deal with the following linear combinations
\begin{eqnarray*}
\frac{1}{4} t_1^{(4)} = \frac{3 C_{22,00}^{22}+\sqrt{3} C_{22,20}^{22}}{12\sqrt{5}}, &&
\frac{1}{4} t_1^{(4)}  x_1^{(4)} =  -\frac{C_{22,20}^{22}}{2\sqrt{15}}, \\ 
t_2^{(4)} = \frac{3 C_{11,00}^{31} + \sqrt{3} C_{11,20}^{31}}{18}, &&
t_2^{(4)}  x_2^{(4)} =  -\frac{ \sqrt{3}C_{11,20}^{31}}{9}, \\
t_e^{(4)} = - \frac{C_{22,22}^{22} }{2\sqrt{105}} , &&
t_o^{(4)} = -\frac{ C_{11,22}^{33}}{30\sqrt{7}}. 
\end{eqnarray*}
We can thus write the fourth order pseudopotential in a Skyrme-like form as
\begin{eqnarray}
\hat V ^{(4)}_{\rm Sk} &=& 
\frac{1}{4} t_1^{(4)} (1+x_1^{(4)} P_{\sigma}) \left[({\bf k}^2 + {\bf k'}^2)^2 + 4 ({\bf k'} \cdot {\bf k})^2\right] \nonumber \\
&+& t_2^{(4)} (1+x_2^{(4)} P_{\sigma}) ({\bf k} \cdot {\bf k'}) ({\bf k}^2 + {\bf k'}^2) \nonumber \\
&+&  t_e^{(4)} \left[  ({\bf k}^2+{\bf k'}^2) T_e({\bf k'},{\bf k})  
+  2 ({\bf k} \cdot {\bf k'}) T_o({\bf k'},{\bf k})  \right] \nonumber \\
&+& t_o^{(4)} \left[ 5 ({\bf k} \cdot    {\bf k'}) T_e({\bf k'},{\bf k})
- \frac{1}{2} ({\bf k}^2+{\bf k'}^2) T_o({\bf k'},{\bf k}) \right] , 
\label{V-four}
\end{eqnarray}
where we have defined two operators involving momenta and spins
\begin{eqnarray}
 T_e({\bf k'},{\bf k}) &=& 3 (\vec \sigma_1 \cdot {\bf k'}) (\vec \sigma_2 \cdot {\bf k'}) 
+ 3 (\vec \sigma_1 \cdot {\bf k}) (\vec \sigma_2 \cdot {\bf k}) 
- ({\bf k'}^2 + {\bf k}^2) (\vec \sigma_1 \cdot \vec \sigma_2), \\
T_o({\bf k'},{\bf k}) &=& 3 (\vec \sigma_1 \cdot {\bf k'}) (\vec \sigma_2 \cdot {\bf k}) 
+ 3 (\vec \sigma_1 \cdot {\bf k}) (\vec \sigma_2 \cdot {\bf k'}) 
- 2 ({\bf k'} \cdot {\bf k}) (\vec \sigma_1 \cdot \vec \sigma_2),
\end{eqnarray} 
which are even and odd under parity transformation, respectively. 
In all these expressions, a $\delta({\bf r}_1-{\bf r}_2)$ function is to be understood, which nevertheless we have omitted for the sake of simplicity. 

The definition of 4th order parameters has been chosen such that their contributions to 
the equation of state and Landau parameters of symmetric nuclear matter maintain a close analogy with those of 2nd order, as shown below. These six parameters are actually the obvious extension to 4th order of the standard $t_{1,2}$, $x_{1,2}$ and $t_{e,o}$ parameters.

Two important remarks are now in order concerning the form of $\hat V ^{(4)}_{\rm Sk}$. 
The first one concerns the so-called D-wave term contained in Skyrme's original proposal \cite{sky59} in the form: 
\begin{equation}
\hat V ^{(D)} = \frac{t_D}{2} \left[ {\bf k'}^2 {\bf k}^2 -  ({\bf k'} \cdot {\bf k})^2  \right] .
\label{d-wave}
\end{equation}
Actually this term is not a pure D-wave but also contains a S-wave contribution. 
It has received some recent attention as a possible improvement of the standard Skyrme interaction \cite{ben13}. It has been wrongly identified in \cite{rai11} as the contribution $\hat{V}_{20,00}^{20}$ (see equation (10) in that article). 
However it does not and can not  appear alone in $\hat{V}^{(4)}_{\rm Sk}$ because it is not gauge invariant. By inspecting (\ref{V-four}) one can see that the bilinear combination related to $t_1^{(4)}$ includes both terms, ${\bf k'}^2 {\bf k}^2$ and $({\bf k'} \cdot {\bf k})^2$, although with different weight each and combined with other powers of momenta. One can also easily verify that the D-wave (\ref{d-wave}) does not contribute to the equation of state of symmetric nuclear matter, neither to the $\ell =2$  Landau parameters, contrarily to the $t_1^{(4)}$ term, whose explicit contribution will be given later on. 

The second remark is the absence of a spin-orbit contribution in the form suggested by Bell and Skyrme \cite{bell56}~:
\begin{equation}\label{so-skyrme} 
i ( \bsigma_1+ \bsigma_2) ({\bf k'} \wedge {\bf k}) F\left({\bf k'}^2, {\bf k}^2, ({\bf k'} \cdot {\bf k}) \right) , 
\end{equation}
where $F$ is a scalar bilinear function of momenta. In the process of constructing the pseudopotential (\ref{hatV}) we have obtained two terms of this type, containing the scalar contributions ${\bf k'}^2 +{\bf k}^2$ and $({\bf k'} \cdot {\bf k})$, which {\em a priori} could be guessed on general grounds. However, neither one of them nor a combination of them fulfill the condition (\ref{gauge}) and thus have been discarded. 
Nevertheless $\hat{V}^{(4)}_{\rm Sk}$ does contain tensor terms which induce spin-orbit effects. Some attempts have been made in the past to explain the spin-orbit coupling in terms of higher order effects of the two-body tensor force \cite{ter60,ari60}. More recently, in a mean-field study of exotic nuclei \cite{otsu06} it has been shown that tensor and two-body spin-orbit interactions produces effects of the same order of magnitude.   
To this respect, it is interesting to observe that the spin-dependent parts proportional to $t_e^{(4)}$ and $t_o^{(4)}$ can actually be rewritten with terms like $(\bsigma_1 . ({\bf k'} \wedge {\bf k}))(\bsigma_2 . ({\bf k'} \wedge {\bf k}))$ which can be interpreted as higher-order spin-orbit contributions \cite{ring80}. It is worth mentioning that the close connection between tensor and spin-orbit forces has also been pointed out in a different physical system, as dipolar Fermi gases \cite{sog12}.

\section{Symmetric nuclear matter properties}\label{inm}

The introduction of a fourth order pseudopotential, or some of its terms, could in principle improve the description of nuclear properties. Considering the recent work about finite size instabilities \cite{pas12a,pas12b,pas12c} one can be worried about the occurrence of such instabilities with an increasing number of terms, in particular with increasing powers of momenta. 
In that sense, the present $\hat{V}^{(4)}_{\rm Sk}$ is not ready  for practical applications as its parameters are not yet fixed nor constrained. 
 At present the only application of the N$^{2}$LO has been presented within the context of the density matrix expansion. 
 A converging expansion  in power of momenta has been obtained in \cite{car10}, thus allowing to convert the interaction energies characteristic to finite and short-range nuclear effective forces into quasi-local density functionals. 
 The parameters obtained in this promising way 
 can be considered as a starting point for a complete minimization procedure~\cite{kor10}.

We now give explicitly the contribution of  $\hat{V}^{(4)}_{\rm Sk}$ to some properties of symmetric nuclear matter namely, the equation of state and the Landau parameters. The contribution to the energy per particle reads
\begin{equation}
\label{contrib}
(E/A)^{(4)} = \frac{9}{280} \left[ 3 t_1^{(4)} + (5 + 4 x_2^{(4)}) t_2^{(4)} \right] \rho k_F^4,
\end{equation}
where $\rho$ is the density and $k_F=(3 \pi^2 \rho /2)^{1/3}$ the Fermi momentum. This term has to be added to the standard equation of state obtained up to second order  given for instance in \cite{cha97}. Notice that the above combination of 4th order parameters is the same as its analogous 2nd order one, apart from a numerical factor and the additional $k_F^2$ power. We thus conclude that $\hat{V}^{(4)}_{\rm Sk}$ modifies the effective mass, introducing an additional density dependence in it. 
In the case of finite nuclei one would expect additional density gradient terms, which can be of fundamental importance to have a peaked value at the surface, which is a crucial element for spectroscopy \cite{ma83}. For completeness we also give the contribution of $\hat{V}^{(4)}_{\rm Sk}$ to the incompressibility at the saturation density $\rho_0$:
\begin{equation}
K_{v}^{(4)}  
=\frac{9}{10} \left[ 3 t_1^{(4)} + (5 + 4 x_2^{(4)}) t_2^{(4)} \right] \left( \frac{3 \pi^2}{2} \right)^{4/3} \rho_0^{7/3}.
\end{equation}

Finite size instabilities in nuclear matter are detected using the linear response, which in general is not an easy task \cite{gar92,dav09,pas13}.  A first insight can nevertheless be obtained in the Landau limit. To this purpose we have calculated the contribution of $\hat{V}^{(4)}_{\rm Sk}$ to the Landau parameters of symmetric nuclear matter, in terms of which the particle-hole interaction is written as \cite{mig67}  
\begin{eqnarray}\label{landau:expr}
\label{landau}
&& 
\sum_{l} \,
\bigg\{ f_{l} + f_{l}' \, (\btau_1 \cdot \btau_2) + \left[ g_{l} 
                 + g_{l}' (\btau_1 \cdot \btau_2 )\right] (\bsigma_1 \cdot \bsigma_2 )  \nonumber \\
&&  \quad 
+ \left[ h_{l}  + h_{l}' (\btau_1 \cdot \btau_2 ) \right]  \frac{k_{12}^{2}}{k_{F}^{2}} \,\, S_{12}
\bigg\} \, P_{l} ( \cos \theta )
\end{eqnarray}
For the tensor parameters we have followed the definition of \cite{dab76,bac79}. The 4th order pseudopotential only contributes to $l \le 2$ central parameters and $\l \le 1$ tensor parameters. We write them in the following way:
\begin{eqnarray}
f_0 &=& \frac{1}{4} L_0[f] + \frac{1}{8} k_F^2 L_2[f] + \frac{1}{6} k_F^4 L_4[f] \label{fgh:param1} \\ 
f_1 &=& - \frac{1}{8} k_F^2 L_2[f] - \frac{1}{4} k_F^4 L_4[f] \label{fgh:param2} \\ 
f_2 &=&  \frac{1}{12} k_F^4 L_4[f] \label{fgh:param3} \\
h_0 &=& \frac{1}{4} k_F^2 L_2[h] + \frac{1}{2} k_F^4 L_4[h] \label{fgh:param4} \\ 
h_1 &=&  - \frac{1}{2} k_F^4 L_4[h]  \label{fgh:param5}
 \end{eqnarray}
and analogously for $f', g, g'$ and $h'$. 
In this notation the power of $k_F$ reflects the contribution coming from each order. 
The explicit expressions of the $L_{n=0,2,4}$ functions entering (\ref{fgh:param1}-\ref{fgh:param5}) are the following. 
At zeroth order we have 
\begin{eqnarray*}
L_0[f] &=& 3 t_0,   \\
L_0[g] &=& -  t_0 (1-2 x_0) ,  \\
L_0[f'] &=&  -  t_0 (1+2 x_0) , \\
L_0[g'] &=& -  t_0, 
 \end{eqnarray*}
 with no density-dependent contribution as previously explained. At second order,
  \begin{eqnarray*}
L_2[f] &=& 3 t_1 + (5+4x_2) t_2, \\ 
L_2[g] &=&  - (1-2x_1) t_1 + (1+2x_2) t_2 , \\ 
L_2[f'] &=& -(1+2x_2) t_1 + (1+2x_2) t_2 ,  \\ 
L_2[g'] &=& - t_1 +  t_2,    \\
L_2[h] &=& t_e + 3 t_o,  \\ 
L_2[h'] &=&  - t_e+t_o.
 \end{eqnarray*}
Finally, at fourth order we have
 \begin{eqnarray*}
L_4[f] &=& 3 t_1^{(4)} + (5+4x_2^{(4)}) t_2^{(4)}, \\ 
L_4[g] &=&  - (1-2x_1^{(4)}) t_1^{(4)} + (1+2x_2^{(4)}) t_2^{(4)},  \\ 
L_4[f'] &=& -(1+2x_2^{(4)}) t_1^{(4)} + (1+2x_2^{(4)}) t_2^{(4)},   \\ 
L_4[g'] &=& - t_1^{(4)} +  t_2^{(4)},    \\
L_4[h] &=&t_e^{(4)} +3t_o^{(4)},  \\ 
L_4[h'] &=&  - t_e^{(4)}  +t_o^{(4)}.
 \end{eqnarray*}
Discarding the 4th order contributions $L_{4}$, these results agree with the ones given in \cite{cao10}  for a standard Skyrme interaction including tensor terms.

For purely central interactions, the stability of the spherical Fermi surface of nuclear matter against small deformations can be expressed in terms of the Landau parameters, which should fulfill some well-known inequalities \cite{mig67}. 
The inclusion of tensor terms produces a coupling between the spin-dependent parts, 
leading to generalized stability criteria, which have been given in a compact form in \cite{bac79} and recently discussed 
for current effective and microscopic interactions in \cite{nav13}. 
 These generalized criteria put additional constraints on the interaction parameters, which could be of great help during the optimization procedure to avoid regions of instabilities.

\section{Conclusions}\label{conclusion}

In summary, we have presented the explicit expression in Cartesian basis of the most general 4th-order contributions to a Skyrme-type pseudopotential  $\hat{V}^{(4)}_{\rm Sk}$ fulfilling gauge invariance. 
Although the requirement of this symmetry is still under debate in the general case, we have chosen to impose it order by order to this particular type of pseudopotential because it ensures the validity of the continuity equation.
Besides, this choice significantly reduces to six the number of independent pseudopotential parameters, which are  
an obvious extension to 4th order of the standard $t_{1,2}$, $x_{1,2}$ and $t_{e,o}$ parameters.

We have observed that $\hat{V}^{(4)}_{\rm Sk}$ contains neither the original D-wave term suggested by Skyrme nor higher order spin-orbit terms. We have used $\hat{V}^{(4)}_{\rm Sk}$ in the context of symmetric nuclear matter  to obtain its explicit contribution to the equation of state and related quantities.
We have also derived the explicit expression of the Landau parameters, interesting in themselves due to their universal character. Indeed, numerical values obtained within some microscopic approach could be used to put explicitly constraints on the values of the coupling constants. 
Moreover, they could be of some help during the minimization procedure of fixing parameters to avoid unstable regions.
Finally, the close connection between higher-order spin-orbit and tensor terms has been pointed out and is believed to hold not only at N$^2$LO but also at N$^3$LO.
Work along these lines is in progress.

\section*{Acknowledgments}
This work was supported by NESQ project (ANR-BLANC 0407, France) and Mineco (Spain), grant FIS2011-28617-C02-02. The authors thank M. Bender, K. Bennaceur, B.G. Carlsson, J. Dobaczewski, T. Duguet, J. Meyer and F. Raimondi for stimulating and encouraging discussions and for useful comments.

\end{document}